\documentclass[reprint,aps,prc,superscriptaddress,amsmath,amssymb]{revtex4-2}
\usepackage{graphicx}
\usepackage{bm}
\usepackage[hidelinks]{hyperref}

\begin{document}

\title{Causal-Horizon Scaling of Quarkonium Suppression in Strong QCD Fields}

\author{Yi Yang}
\email{yiyang429@as.edu.tw}
\affiliation{Institute of Physics, Academia Sinica, Taipei 11529, Taiwan}
\affiliation{Department of Physics, National Cheng Kung University, Tainan 70101, Taiwan}

\begin{abstract}
The simultaneous observation of strong sequential bottomonium suppression and small azimuthal anisotropy constrains the time scale and geometry of quarkonium dissociation in relativistic heavy-ion collisions. This work investigates an early-time contribution in which strong pre-equilibrium color fields generate an effective proper-acceleration scale and an associated causal length. The maximal acceleration is fixed through the phenomenological anchoring assumption $T_U^{\max}\simeq T_c$, without identifying the kinematic Unruh scale with an equilibrium temperature. The survival probability is described by an event-averaged one-scale exponential ansatz. With no state-by-state adjustment, the resulting horizon component gives a quantitatively reasonable description of the LHC centrality dependence. For the directly measured $\Upsilon(2S)$-to-$\Upsilon(1S)$ double ratio, the fixed horizon term supplies a substantial part of the relative suppression, while the data require an additional state-dependent late-stage factor. A minimal one-parameter effective QGP attenuation gives a quantitative multistage description. At RHIC, the absolute $R_{AA}$ central values lie below the isolated horizon component, while the relative $2S$-to-$1S$ suppression remains compatible with the predicted state-size hierarchy within present uncertainties. Because the proposed early factor is local and scalar, it gives the exact leading-order null contribution $v_2^H=0$ within this construction, consistent with current CMS measurements.
\end{abstract}

\maketitle

\section{Introduction}
\label{sec:intro}

Heavy quarkonia are among the most extensively studied probes of strongly interacting matter in relativistic nuclear collisions. Because the heavy-quark mass provides a hard production scale while the subsequent bound-state evolution is sensitive to the surrounding color environment, quarkonium observables connect initial hard production, pre-equilibrium dynamics, and the later quark-gluon plasma (QGP). The sequential ordering of bottomonium nuclear modification factors,
$R_{AA}(1S)>R_{AA}(2S)>R_{AA}(3S)$, reflects the different binding energies and spatial sizes of the states and has long been used to constrain in-medium dissociation mechanisms \cite{Matsui:1986dk,Rothkopf:2019ipj,Andronic:2015wma,Brambilla:2010cs}.

Precision measurements by CMS \cite{CMS:Upsilon,CMS:v2}, ALICE \cite{ALICE:v2}, ATLAS \cite{ATLAS:Jpsi}, and STAR \cite{STAR:2022sequential} now provide centrality-, momentum-, and collision-energy-dependent information. The measured bottomonium elliptic flow is small within present uncertainties, while the suppression is strong and sequential. These two observations jointly constrain how much suppression is generated locally at early time and how much is accumulated along trajectories through the later anisotropic medium.

The contemporary theoretical landscape contains several complementary approaches. Transport calculations combine dissociation, feed-down, regeneration, and hydrodynamic space-time evolution \cite{Rapp:2017chg,Du:2017qkv}. Potential-based and real-time quantum-evolution frameworks follow the in-medium heavy-quark wave function, including finite-temperature screening and dynamical decoherence \cite{Rothkopf:2019ipj,Strickland:2011mw,Islam:2020,HQQDfluct}. Open-quantum-system (OQS) formulations describe the reduced density matrix of the heavy pair and systematically organize singlet--octet transitions and medium-induced noise \cite{Brambilla:2020qwo,Akamatsu:2020ypb,Yao:2020eqy,StricklandThapa:2023}. These modern treatments can reproduce important features of bottomonium suppression and can also yield very small $v_2$. The near-zero anisotropy therefore does not by itself indicate a failure of conventional QGP descriptions.

Nevertheless, the present data do not uniquely determine the time ordering of the suppression. A small final-state $v_2$ is compatible with a late-time mechanism whose anisotropy remains weak, but it is also compatible with a non-negligible local suppression component established before substantial transverse hydrodynamic response develops. This motivates asking whether strong pre-equilibrium color fields can supply an additional geometric survival factor that precedes the conventional QGP evolution.

\begin{figure*}[t]
 \centering
 \includegraphics[width=0.96\textwidth]{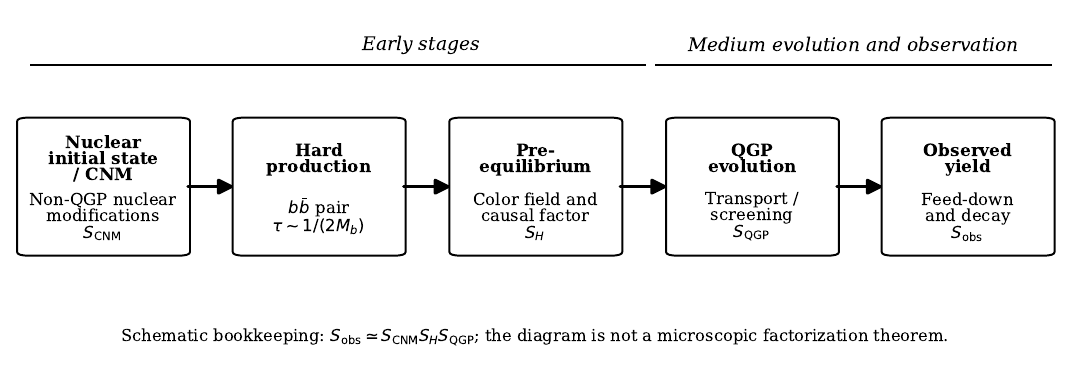}
 \caption{Schematic organization of the multistage bottomonium evolution. Here $S_{\rm CNM}$ denotes cold-nuclear-matter modifications associated with the nuclear initial state, $S_H$ is the proposed pre-equilibrium causal-horizon factor, and $S_{\rm QGP}$ denotes subsequent in-medium evolution. Feed-down modifies the final inclusive yields. The diagram is conceptual and does not imply an exact microscopic factorization or a strict temporal separation of all nuclear effects.}
 \label{fig:timeline}
\end{figure*}

This work examines such a possibility through a causal scale associated with strong-field acceleration \cite{Castorina:2007eb,Hawking:1975}. The objective is not to replace transport or OQS calculations, nor to claim that an exponential functional form uniquely identifies a causal mechanism. Rather, the analysis tests whether one early geometric scale can simultaneously account for the LHC state hierarchy, the centrality-dependent double ratio, the weak $p_T$ dependence, and the absence of an additional elliptic-flow component. Figure~\ref{fig:timeline} summarizes how this contribution is embedded in a multistage picture.

\section{Effective acceleration and causal scale}
\label{sec:acceleration}

Heavy quark-antiquark pairs are produced through hard partonic scatterings on time scales $\tau\sim1/(2M_b)\ll0.1$ fm/$c$. Immediately after production, the pair is embedded in rapidly evolving longitudinal color fields characteristic of the Glasma stage \cite{Lappi:2006fp,Gelis:2010nm}. In a simplified local picture, the force acting on a heavy color charge defines a proper-acceleration scale of order $\sigma_{\rm eff}/M_Q$, where $\sigma_{\rm eff}$ denotes an effective field or string-tension scale. The microscopic field is neither homogeneous nor stationary, so this relation should not be interpreted as an eternally constant acceleration.

An early-time effective acceleration is therefore introduced,
\begin{equation}
 \bar a(\bm x_\perp)=
 \frac{\displaystyle\int_0^{\Delta\tau}d\tau\,w(\tau)
 \frac{\sigma_{\rm eff}(\tau,\bm x_\perp)}{M_Q}}
 {\displaystyle\int_0^{\Delta\tau}d\tau\,w(\tau)},
 \label{eq:aeff}
\end{equation}
where $w(\tau)$ represents the temporal sensitivity of the pre-hadronic pair. Equation~(\ref{eq:aeff}) is a coarse-grained definition of the scale entering the phenomenological model. A first-principles evaluation of $w(\tau)$ and $\sigma_{\rm eff}(\tau,\bm x_\perp)$ would require real-time classical-field and heavy-pair evolution and is beyond the present proof-of-concept treatment.

The corresponding causal length is
\begin{equation}
 r_H=\bar a^{-1}.
 \label{eq:rH}
\end{equation}
For comparison with standard acceleration notation, one may introduce $T_U=\bar a/(2\pi)$ and write $r_H=(2\pi T_U)^{-1}$ \cite{Unruh:1976}. In the nonstationary environment considered here, $T_U$ is only a kinematic acceleration scale. It is neither identified with the thermodynamic temperature of an equilibrated QGP nor taken to imply that the pair observes an exact stationary Unruh spectrum. The use of this notation is intended to emphasize the causal length associated with a large proper acceleration.

\section{Event-averaged survival ansatz}
\label{sec:survival}

For a bound state to form, the color correlation required across its characteristic spatial extent must remain dynamically connected. A literal classical implementation in a single event could be represented by a threshold $\Theta(r_H-r_{nS})$. Such a step function, however, treats both the causal boundary and the bound-state size as sharp quantities. In practice, $r_{nS}$ is an rms scale of an extended quantum state, while the effective causal length varies with transverse position, proper time, and event-by-event field fluctuations.

The experimentally relevant survival factor is therefore modeled as an ensemble average,
\begin{equation}
 S_{nS}^{\rm raw}=\int_0^\infty dr_H\,P(r_H)\,\Theta(r_H-r_{nS}).
 \label{eq:ensemble}
\end{equation}
To retain a minimal one-scale construction, the model assumes the positive distribution
\begin{equation}
 P(r_H)=\frac{1}{\bar r_H}\exp\left(-\frac{r_H}{\bar r_H}\right),
 \label{eq:phorizon}
\end{equation}
which gives
\begin{equation}
 S_{nS}^{\rm raw}=\exp\left(-\frac{r_{nS}}{\bar r_H}\right)
       =\exp\left(-\bar a\,r_{nS}\right).
 \label{eq:survivalraw}
\end{equation}
This construction makes explicit why an ensemble-averaged quantum survival probability need not vanish abruptly when a characteristic rms radius exceeds the mean causal length. It remains a phenomenological one-scale realization rather than a unique microscopic derivation. Other fluctuation distributions, explicit wave-function overlaps, or real-time quantum evolution would test the robustness of the functional form.

The experimentally relevant nuclear modification is normalized to the $pp$ reference. Here $\bar a_{AA}\equiv\bar a(N_{\rm part},\sqrt{s_{NN}})$ denotes the ensemble-averaged effective acceleration scale for the selected nucleus--nucleus centrality class, while $\bar a_{pp}\equiv\bar a(N_{pp},\sqrt{s_{NN}})$ denotes the corresponding reference scale evaluated for $pp$ collisions at the same energy, with $N_{pp}=2$. The horizon contribution therefore enters as
\begin{equation}
 S_H^{nS}=\frac{S_{nS}^{AA}}{S_{nS}^{pp}}
 =\exp\left[-r_{nS}\left(\bar a_{AA}-\bar a_{pp}\right)\right].
 \label{eq:survival}
\end{equation}
A similar exponential dependence can arise in a reduced description of color screening if the causal length is replaced by an in-medium screening length. The exponential form is therefore not mechanism-specific. The physical hypothesis tested here concerns the origin and space-time evolution of the controlling scale: in the present framework it is established locally during the pre-hydrodynamic stage, whereas conventional in-medium suppression evolves with the subsequent QGP background.

\section{Centrality and collision-energy scaling}
\label{sec:scaling}

To connect the local causal scale with collision geometry, a mean-field centrality parametrization is adopted. Glauber geometry provides the standard relation between centrality, $N_{\rm part}$, and the characteristic nuclear overlap \cite{Miller:2007ri}. Motivated by the growth of a nuclear linear scale and average thickness, the cube-root proxy is used:
\begin{equation}
 \bar a(N_{\rm part},\sqrt{s_{NN}})=
 \kappa_{\rm LHC}
 \left(\frac{\sqrt{s_{NN}}}{\sqrt{s_{\rm LHC}}}\right)^\lambda
 N_{\rm part}^{1/3}.
 \label{eq:scaling}
\end{equation}
Equation~(\ref{eq:scaling}) is a phenomenological geometric ansatz, not an exact consequence of the Glauber model. Its energy dependence is motivated by small-$x$ saturation phenomenology, with $\lambda\in[0.2,0.3]$ adopted here \cite{Kharzeev:2000ph,Gelis:2010nm}. Because related initial-state scalings are also used in hydrodynamic and transport calculations, neither power is exclusive to the causal interpretation.

The overall scale is fixed by the explicit phenomenological anchoring condition
\begin{equation}
 T_U^{\max}=\frac{\bar a(N_{\rm part,max},\sqrt{s_{\rm LHC}})}{2\pi}
 \simeq T_c,
 \label{eq:anchor}
\end{equation}
for central Pb+Pb collisions. This is a model postulate that relates the maximal acceleration scale to the characteristic QCD deconfinement scale; it is not a thermodynamic identity. For $T_c\simeq140$--150 MeV and $N_{\rm part,max}\simeq380$, Eq.~(\ref{eq:anchor}) gives $\kappa_{\rm LHC}\simeq0.62$--0.66 fm$^{-1}$; the representative value $0.63$ fm$^{-1}$ is used. Thus the normalization is fixed before comparison with individual bottomonium states.

The $pp$-normalized attenuation scale in Eq.~(\ref{eq:survival}) is
\begin{equation}
 \Delta\bar a=
 \kappa_{\rm LHC}
 \left(\frac{\sqrt{s_{NN}}}{\sqrt{s_{\rm LHC}}}\right)^\lambda
 \left(N_{\rm part}^{1/3}-N_{pp}^{1/3}\right),
 \qquad N_{pp}=2.
 \label{eq:deltascale}
\end{equation}
The calculation uses the representative potential-model radii $r_{1S}=0.28$ fm, $r_{2S}=0.56$ fm, and $r_{3S}=0.78$ fm \cite{Karsch:1988gun}. These are phenomenological size inputs rather than model-independent experimental radii.

\begin{figure}[t]
 \centering
 \includegraphics[width=\linewidth]{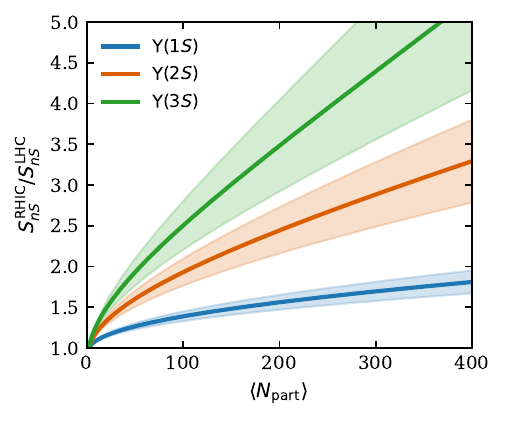}
 \caption{Horizon-baseline expectation for the ratio of survival probabilities between RHIC (200 GeV) and the LHC (5.02 TeV). The bands correspond to $\lambda\in[0.2,0.3]$.}
 \label{fig:energy}
\end{figure}

The resulting leading energy-scaling expectation is
\begin{align}
 \frac{S_{nS}^{\rm RHIC}(N_{\rm part})}{S_{nS}^{\rm LHC}(N_{\rm part})}
 =\exp\Bigg[&r_{nS}\kappa_{\rm LHC}
 \left(N_{\rm part}^{1/3}-N_{pp}^{1/3}\right)\nonumber\\
 &\times\left\{1-
 \left(\frac{\sqrt{s_{\rm RHIC}}}{\sqrt{s_{\rm LHC}}}\right)^\lambda\right\}\Bigg].
 \label{eq:ratio}
\end{align}
Figure~\ref{fig:energy} shows this one-scale expectation. It is a falsifiable consequence of the assumed energy dependence, although not a mechanism-exclusive signature.

\section{Phenomenological comparison and multistage scope}
\label{sec:phenom}

In a realistic collision, quarkonium production and survival receive contributions from several stages. Cold-nuclear-matter (CNM) modifications affect the nuclear initial state and pre-resonance propagation; feed-down changes the inclusive yields; and the later QGP produces screening, dissociation, decoherence, and possibly regeneration. As schematic bookkeeping in a separated-stage limit, the contributions can be written as
\begin{equation}
 S_{\rm obs}^{nS}\simeq
 S_{\rm CNM}^{nS}\,S_H^{nS}\,S_{\rm QGP}^{nS}.
 \label{eq:factorization}
\end{equation}
Equation~(\ref{eq:factorization}) is not a precision factorization theorem. Its purpose is to identify the proposed $S_H$ as an early component without excluding established CNM or QGP physics.

\begin{figure*}[t]
 \centering
 \includegraphics[width=0.98\textwidth]{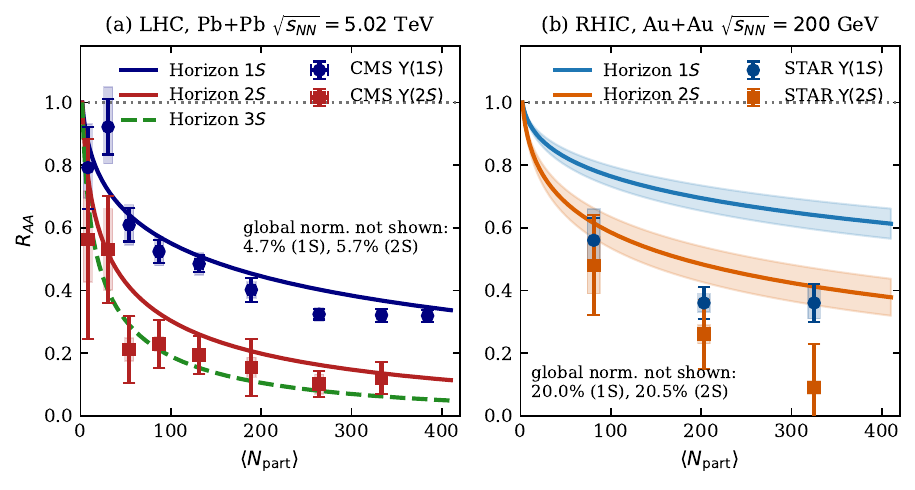}
 \caption{Causal-horizon component compared with official (a) CMS Pb+Pb data at 5.02 TeV and (b) STAR Au+Au data at 200 GeV. Bars denote statistical uncertainties and boxes point-to-point systematic uncertainties. Global normalization uncertainties are quoted in the panels and are not shown on each point. No additional CNM or QGP correction is applied to the curves.}
 \label{fig:raa}
\end{figure*}

Figure~\ref{fig:raa} compares the causal-horizon component alone with the official CMS and STAR measurements. Statistical uncertainties are shown by bars and point-to-point systematic uncertainties by boxes; the quoted global normalizations are listed in the panels but not folded into the points. At the LHC, the same scale gives the observed ordering and a quantitatively reasonable centrality dependence. Using statistical and point-to-point systematic errors in quadrature only as an indicative diagnostic gives $\chi^2/\mathrm{ndf}=16.7/9$ for $\Upsilon(1S)$ and $7.8/8$ for $\Upsilon(2S)$, without fitting either state. The agreement, particularly for $2S$, shows that the one-scale state-size dependence is nontrivial; it does not imply that CNM, feed-down, or QGP effects vanish, since different mechanisms can be degenerate in inclusive $R_{AA}$.

At RHIC, the absolute horizon-only curves lie above the STAR central values, especially for $1S$. However, the $2S$-to-$1S$ suppression ratio is more compatible with the data. For the 0--60\% integrated values, STAR reports $R_{AA}(1S)=0.40$ and $R_{AA}(2S)=0.26$, giving a central ratio of $0.65$, while the horizon baseline predicts $0.68$--$0.75$ for $\lambda=0.2$--0.3. Within the current uncertainties, the main RHIC difference is therefore compatible with an additional approximately state-common suppression component rather than a failure of the predicted size ordering. No stronger feed-down contribution at RHIC is invoked. The interpretation must also account for the sizable normalization uncertainty of about 20\% and for the broad uncertainty bands of current transport and OQS calculations, arising for example from CNM inputs and transport coefficients; STAR finds those calculations compatible with the data even though their central $1S$ predictions tend to lie above the measurements \cite{STAR:2022sequential}.

The mathematical similarity between Eq.~(\ref{eq:survival}) and simplified screening or absorption models is acknowledged. Likewise, the cube-root centrality dependence and CGC-inspired energy scaling do not uniquely identify a causal mechanism. The proposed distinction is one of space-time origin: $S_H$ is established locally before substantial hydrodynamic transverse response, whereas transport and OQS descriptions accumulate an important part of the suppression during QGP evolution. The two contributions can coexist, and the double-ratio analysis below makes the additional state-dependent late-stage contribution explicit rather than setting it to unity.

\section{Relative state suppression and the double ratio}
\label{sec:double}

The directly measured double ratio
\begin{equation}
 D_{21}^{\rm obs}=\frac{[\Upsilon(2S)/\Upsilon(1S)]_{\rm PbPb}}
 {[\Upsilon(2S)/\Upsilon(1S)]_{pp}}
 \label{eq:doubledef}
\end{equation}
is especially useful because luminosity, $N_{\rm coll}$, the overall $pp$ production normalization, and many common acceptance and efficiency effects cancel. State-common CNM and medium-normalization effects are also strongly reduced. The state-dependent part of QGP suppression, however, does not cancel and must be retained.

The fixed early-time contribution is
\begin{equation}
 D_{21}^{H}=\frac{S_H^{2S}}{S_H^{1S}}
 =\exp\left[-(r_{2S}-r_{1S})\Delta\bar a\right].
 \label{eq:doubleratio}
\end{equation}
It is controlled directly by the radius difference and contains no additional state normalization. In a multistage description the observed ratio is instead
\begin{equation}
 D_{21}^{\rm obs}\simeq D_{21}^{H}
 D_{21}^{\rm CNM}D_{21}^{\rm QGP}.
 \label{eq:doublefactor}
\end{equation}
Because most common CNM effects cancel in this ratio, a minimal effective late-stage factor is used:
\begin{equation}
 D_{21}^{\rm QGP,eff}=
 \exp[-\gamma_{\rm QGP}^{\rm eff}G(N_{\rm part})],
 \qquad
 G=N_{\rm part}^{1/3}-N_{pp}^{1/3},
 \label{eq:qgpeff}
\end{equation}
to quantify the additional state-dependent attenuation required by the data. This is not a microscopic QGP calculation; $\gamma_{\rm QGP}^{\rm eff}$ may also absorb residual non-canceling CNM and feed-down effects. Its purpose is to avoid incorrectly identifying the horizon-only curve with the complete double ratio.

\begin{figure*}[t]
 \centering
 \includegraphics[width=0.98\textwidth]{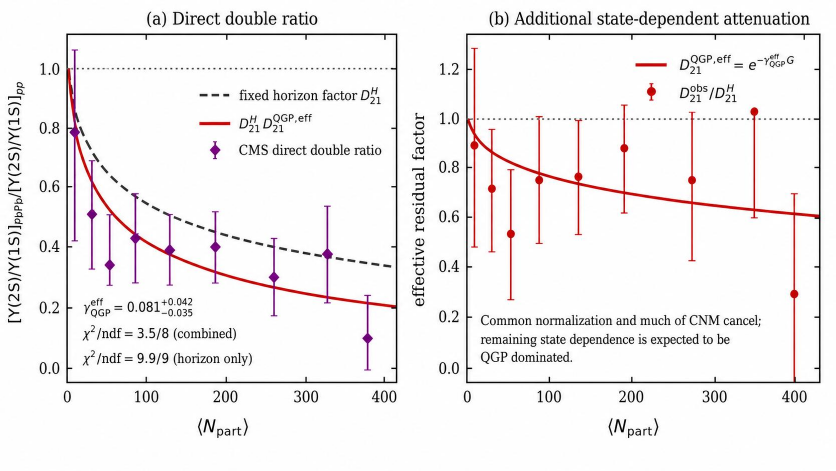}
 \caption{(a) Direct CMS $\Upsilon(2S)$-to-$\Upsilon(1S)$ double ratio \cite{CMS:DoubleRatio}. The dashed curve is the fixed horizon contribution, Eq.~(\ref{eq:doubleratio}); the solid curve includes the fitted effective late-stage factor in Eq.~(\ref{eq:qgpeff}). (b) Residual factor obtained by dividing the measured double ratio by the fixed horizon contribution. Bars and boxes denote statistical and point-to-point systematic uncertainties, respectively. The effective QGP factor is a reduced phenomenological description, not a replacement for transport or OQS calculations.}
 \label{fig:double}
\end{figure*}

Figure~\ref{fig:double} uses the directly measured CMS values rather than a ratio reconstructed from independent $R_{AA}$ tables. The fixed horizon curve already produces a substantial part of the observed decrease but lies systematically above several central values. Fitting the single residual coefficient gives
\begin{equation}
 \gamma_{\rm QGP}^{\rm eff}=0.081^{+0.042}_{-0.035},
 \label{eq:gammafit}
\end{equation}
with $\chi^2/\mathrm{ndf}=3.5/8$, using statistical and point-to-point systematic uncertainties and excluding the common 3.1\% $pp$ uncertainty. The fixed horizon coefficient in the exponent is $\kappa_{\rm LHC}(r_{2S}-r_{1S})=0.176$, so the fitted late-stage coefficient is about 46\% of the early coefficient in this reduced exponential decomposition. The horizon component therefore remains the larger contribution to the centrality-dependent exponent, while an additional state-dependent QGP effect is quantitatively required.

\section{Momentum dependence, elliptic flow, and geometric tests}
\label{sec:pt}

\begin{figure*}[t]
 \centering
 \includegraphics[width=0.98\textwidth]{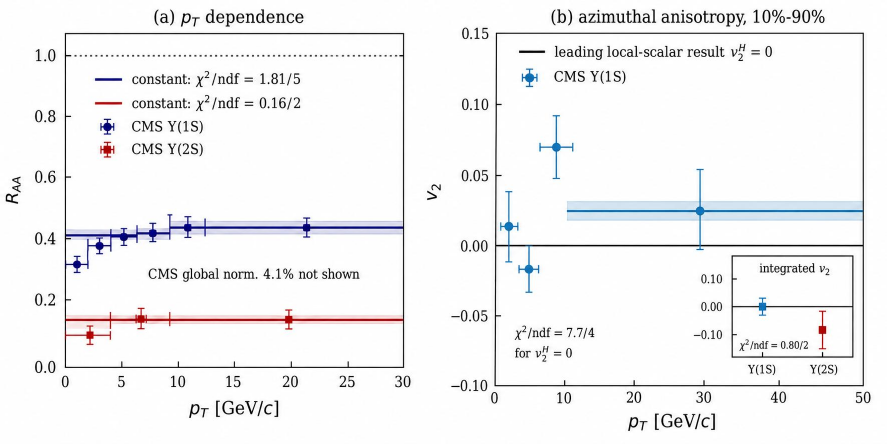}
 \caption{(a) CMS $R_{AA}$ versus $p_T$ with constant shape references. (b) CMS $\Upsilon(1S)$ elliptic flow versus $p_T$ compared with the exact leading local-scalar result $v_2^H=0$; the inset shows the integrated $1S$ and $2S$ measurements. Bars denote statistical uncertainties and boxes systematic uncertainties.}
 \label{fig:ptv2}
\end{figure*}

A realistic production spectrum does not, by itself, generate a $p_T$ dependence in a strictly factorized scalar survival model. If
\begin{equation}
 \frac{dN_{AA}}{dp_T}=N_{\rm coll}\,S_H(p_T)\frac{dN_{pp}}{dp_T},
 \label{eq:yieldpt}
\end{equation}
then
\begin{equation}
 R_{AA}(p_T)=S_H(p_T).
 \label{eq:raapt}
\end{equation}
Consequently, multiplication by the same $pp$ spectrum cancels in the ratio when $S_H$ is momentum independent. Figure~\ref{fig:ptv2}(a) shows the CMS measurements with both statistical and systematic uncertainties. Constant fits give $\chi^2/\mathrm{ndf}=1.81/5$ and $0.16/2$ for $1S$ and $2S$, respectively. This supports a weak $p_T$ dependence, although it does not fix the absolute normalization of the causal component. Formation time, feed-down, CNM effects, production-position--momentum correlations, and QGP trajectories can all generate additional momentum dependence in a complete calculation.

The local horizon factor is scalar with respect to the transverse production direction and therefore generates no additional elliptic anisotropy at leading order,
\begin{equation}
 v_2^H=0.
 \label{eq:v2zero}
\end{equation}
Within the idealized local-scalar construction this is an exact structural null result, not a fitted quantity, and no intrinsic parameter uncertainty is assigned to it. Nonzero corrections can arise once correlations between the fluctuating initial field, production geometry, event plane, and subsequent QGP evolution are included; such effects belong to a more complete calculation rather than to the leading horizon factor itself.

Figure~\ref{fig:ptv2}(b) compares the null contribution with the CMS $\Upsilon(1S)$ $p_T$-differential measurement and also displays the integrated $1S$ and $2S$ results. Combining statistical and point-to-point systematic uncertainties in quadrature gives $\chi^2/\mathrm{ndf}=7.70/4$ for the four differential $1S$ points relative to $v_2^H=0$. For the integrated values $v_2(1S)=0.007\pm0.011\pm0.005$ and $v_2(2S)=-0.063\pm0.085\pm0.037$, the corresponding null test gives $\chi^2/\mathrm{ndf}=0.80/2$ \cite{CMS:v2}. The $2S$ result is available only in integrated form, which is why it is shown in the inset rather than as a $p_T$-differential series. Equation~(\ref{eq:v2zero}) is not a complete prediction for the final observed flow: later transport and OQS dynamics can also yield a small $v_2$. It states only that the proposed early scalar factor does not itself add an anisotropic component.

Because reduced centrality scaling is shared by several mechanisms, more differential tests should condition on event geometry. At fixed centrality or multiplicity, event-shape engineering changes eccentricity and path-length distributions while keeping the overall system size approximately controlled. A predominantly local early-time component should be less sensitive to these later path variations than suppression accumulated during propagation. Initial field fluctuations are themselves correlated with final event shape, so a quantitative event-shape calculation is required before this becomes a decisive discriminator. Rapidity- and energy-differential double ratios provide complementary constraints.

\section{Conclusion}
\label{sec:conclusion}

This work develops a one-scale early-time contribution to bottomonium suppression in which strong pre-equilibrium color fields define an effective acceleration and causal length. A sharp event-level causal condition, averaged over a fluctuating horizon scale, leads to the exponential survival ansatz. The overall normalization is fixed by the phenomenological postulate $T_U^{\max}\simeq T_c$, while the bottomonium-state dependence is determined by representative vacuum radii.

The present framework makes three concrete phenomenological statements. First, the same scale gives the LHC sequential hierarchy and a quantitatively reasonable centrality dependence without state-by-state tuning. Second, the directly measured $2S$-to-$1S$ double ratio cleanly separates much of the common normalization while retaining state-dependent QGP suppression. The fixed horizon contribution supplies the larger part of the centrality-dependent exponent in a reduced multistage decomposition, and one additional effective QGP coefficient brings the double ratio into quantitative agreement. Third, the local scalar factor produces no additional elliptic anisotropy and is consistent with the observed small $v_2$, while the measured $p_T$ dependence is compatible with a constant survival shape.

The RHIC absolute $R_{AA}$ values lie below the isolated horizon baseline, but their relative $2S$-to-$1S$ suppression remains compatible with the predicted hierarchy within current uncertainties. This indicates that an additional approximately state-common contribution may coexist with the early geometric factor; it does not require an enhanced feed-down fraction at RHIC. A complete extraction will require CNM, feed-down, and QGP evolution to be treated together with the pre-equilibrium stage. More precise RHIC data, direct double ratios at several energies, and event-shape-differential measurements can determine whether the early causal component is non-negligible.

\begin{acknowledgments}
The author is grateful for the support provided by the Institute of Physics, Academia Sinica, and the Department of Physics, National Cheng Kung University. This work was supported in part by the National Science and Technology Council of Taiwan.

\paragraph*{Declaration of AI-assisted editing} The author used AI-based language tools for editorial assistance, including readability, wording, and organization. All scientific content, calculations, figures, interpretations, and conclusions were developed, checked, and approved by the author.
\end{acknowledgments}

\section*{Data Availability}
The numerical experimental data used in Figs.~3--5 are publicly available through the HEPData records associated with Refs.~\cite{CMS:Upsilon,CMS:v2,STAR:2022sequential,CMS:DoubleRatio}. The numerical results of the causal-horizon model can be reproduced from the equations and parameter values provided in the manuscript. No additional data or specialized software are required to reproduce the theoretical results.

\end{document}